\documentclass{article}
\def\subparagraph{} 
\usepackage{spconf,amsmath,graphicx,enumitem,}
\usepackage{mathrsfs}
\usepackage{multirow}
\usepackage[flushleft]{threeparttable}
\usepackage[fontsize=9pt]{fontsize}
\usepackage{enumitem}

\AtBeginDocument{%
  \setlength\abovedisplayskip{0pt}
  \setlength\belowdisplayskip{0pt}}

\title{Self-supervised speech representation learning for keyword-spotting with light-weight transformers}

\name{Chenyang Gao$^{\star}$ \thanks{Chenyang Gao performed this work while interning at Amazon.} \qquad Yue Gu$^{\dagger}$ \qquad Francesco Caliva$^{\dagger}$ \qquad Yuzong Liu$^{\dagger}$} 
\address{$^{\star}$Rutgers, The State University of New Jersey \\ $^{\dagger}$Alexa Perceptual Technologies, Amazon}

\begin{document}
\maketitle
\vspace{-2.5cm}
\begin{abstract}
Self-supervised speech representation learning (S3RL) is revolutionizing the way we leverage the ever-growing availability of data. While S3RL related studies typically use large models, we employ light-weight networks to comply with tight memory of compute-constrained devices. We demonstrate the effectiveness of S3RL on a keyword-spotting (KS) problem by using transformers with 330k parameters and propose a mechanism to enhance utterance-wise distinction, which proves crucial for improving performance on classification tasks. On the Google speech commands v2 dataset, the proposed method applied to the Auto-Regressive Predictive Coding S3RL led to a 1.2\% accuracy improvement compared to training from scratch. On an in-house KS dataset with four different keywords, it provided 6\% to 23.7\% relative false accept improvement at fixed false reject rate. We argue this demonstrates the applicability of S3RL approaches to light-weight models for KS and confirms S3RL is a powerful alternative to traditional supervised learning for resource-constrained applications.
\end{abstract}
\vspace{-1.5mm}
\section{Introduction}
\label{sec:intro}
\vspace{-1mm}
With the prosperous growth of deep learning, scarcity of annotated data is considered the bottleneck of model training. The general consensus is that abundant amounts of data favor model generalization. However, collecting large amounts of labeled data is time consuming and error-prone. Hence, the interest in developing methods to effectively leverage unannotated data has increased. Self-supervised learning (SSL) is a widely adopted solution in Natural Language Processing (NLP)~\cite{devlin2018bert, peters-etal-2018-deep} and Computer Vision (CV)~\cite{he2020momentum,he2022masked}. In speech domain, non-annotated data are commonly used to pre-train models based on self-supervised speech representation learning (S3RL)~\cite{yang2021superb}. S3RL methods could be categorized into generative loss-based~\cite{chung2020generative,jiang2019improving,liu2020mockingjay}, latent unit prediction-based~\cite{hsu2021hubert,chang2022distilhubert}, discriminative learning-based approaches~\cite{oord2018representation,ling2020deep,baevski2020wav2vec} or combinations thereof~\cite{ling2020decoar}. These methods are capable of learning meaningful representation of speech. They can use the gained knowledge for instance to infer missing features given a context or to classify data even when relevant features of the input have been masked out. Learned representations from S3RL have been shown to generalize to keyword spotting~\cite{yang2021superb,gong2022ssast} but limitations still exist. S3RL methods require training models with millions of parameters to be able to generate good quality speech representation and this requirement cannot be accommodated for models deployed on mobile and edge devices. Furthermore, S3RL methods have mostly been designed for automatic speech recognition (ASR) applications and have been neglecting the need for learning different representations among utterances (or utterance-wise differences), which is critical for classification tasks such as keyword-spotting (KS), sentiment analysis and speaker identification. To overcome the aforementioned limitations, we employed S3RL on a 330K parameters transformer for KS on-edge. We studied Auto-Regressive Predictive Coding (APC), Masked Predictive Coding (MPC) and Contrastive Learning (CL). A concurrent work used SSL for KS~\cite{bovbjerg2022improving}; however, their approach is based on Data2Vec~\cite{baevski2022data2vec}, which is a general SSL method. Instead, we focused on methods specific to learning speech related representations and used a transformer with half of the number of parameters of the smallest model used in \cite{bovbjerg2022improving}. After pre-training, we added a linear classifier to the light-weight transformer and fine-tuned it to a KS down-stream task. Furthermore, we introduced a novel two-step contrastive-learning mechanism to facilitate learning more diverse utterance-level representations. Our proposed approach can be easily plugged into any S3RL method. We evaluated performance on the publicly available Google speech commands v2 dataset~\cite{warden2018speech} as well as in an in-house KS dataset. Results showed that the proposed approach, when applied to APC S3RL achieved 1.2\% accuracy improvement compared to training from scratch on Google Commands V2 35 classes classification and 6\% to 23.7\% relative false accept improvements at fixed false reject rate on four keywords, in our in-house KS dataset. In summary, our contributions can be summarized as follows:
\begin{itemize}[noitemsep,topsep=0pt]
  \item We explored S3RL in light-weight transformers and demonstrated S3RL effectiveness on keyword-spotting;
  \item We devised a method which can be easily applied to any S3RL method to improve utterance-wise distinction. We show this leads to better keyword-spotting performance;
  \item To the best of our knowledge, we are the first to conduct an extensive and fair comparison among S3RL methods on a common benchmark, drawing conclusion on which S3RL approach is best suited for pre-training light-weight models.
\end{itemize}
\begin{figure*}[!tb]
  \begin{minipage}[tb]{1.0\linewidth}
  \centering
  \centerline{\includegraphics[width=\columnwidth]{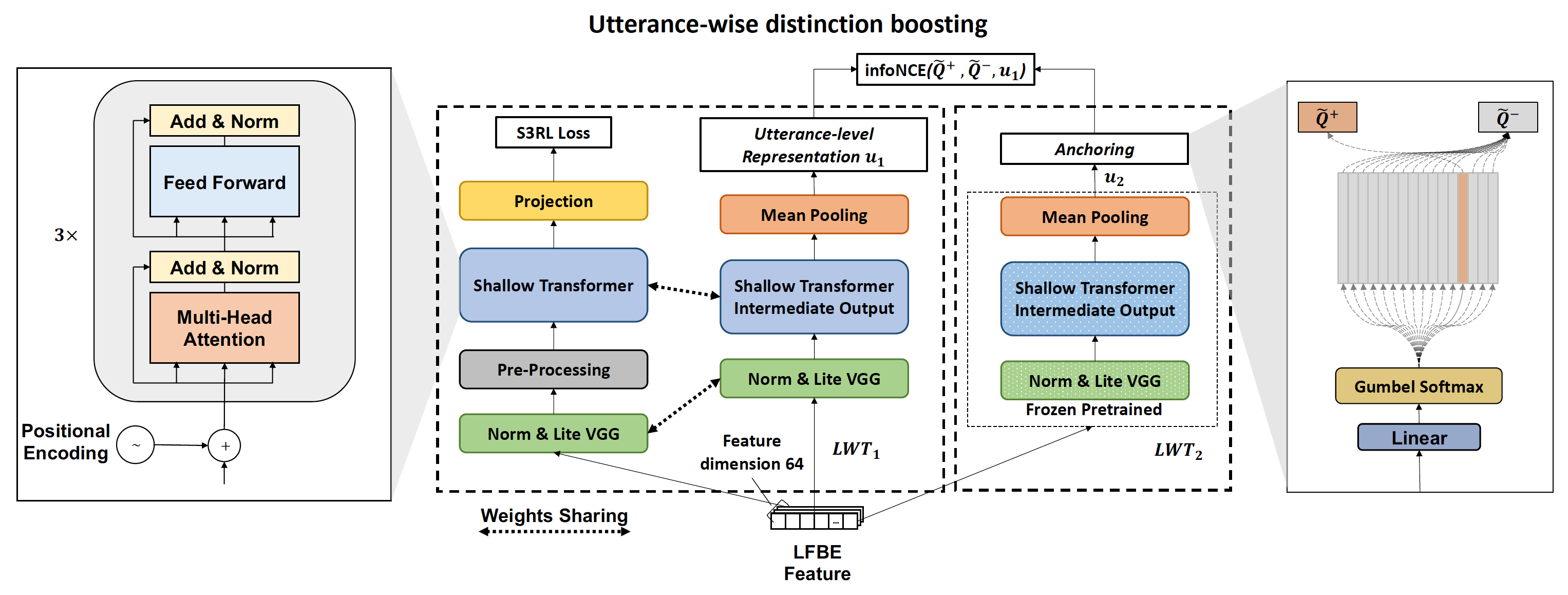}}
  \vspace{-2.5mm}
\caption{Proposed two steps utterance-wise contrastive-learning. The S3RL block performs S3RL on a task. The utterance-wise distinction boosting block helps locating the anchor class for utterance level representation using utterance-wise boosting, which leverages the pre-trained S3RL model.}
\end{minipage}
\vspace{-2.5mm}
\label{figure utterance-wise distinction boosting}
\end{figure*}
\vspace{-1.5mm}
\section{Background}
\label{ssec:s3rl}
\vspace{-1mm}
S3RL has proven effective in pre-training large models for speech-related tasks. We focus on three of these methods.
\textbf{Auto-regressive Predictive Coding}~\cite{chung2020generative} is a generative-loss based SR3L approach, which allows training auto-regressive models, meaning that by using information from past timestamps they can predict future information. 
For this, APC relies on the training objective function reported in eq.~\ref{APC}, where $\mathbf{X}$=($x_{1}$, $x_{2}$,...,$x_{T}$) is the target sequence, $\mathbf{Y}$=($y_{1}$, $y_{2}$,...,$y_{T}$) is the predicted sequence and $T$ represents the total sequence length. 
APC predicts a future frame $n$ steps ahead of the current frame.
\begin{equation}\label{APC}
   \mathscr{L}_{APC} = \sum^{T-n}_{i=1} \left \| x_{i+n} - y_{i} \right \|
\end{equation}
\textbf{Masked Predictive Coding}~\cite{jiang2019improving,liu2020mockingjay} is another generative-loss based SR3L approach. It uses bi-directional context to reconstruct masked features. MPC is generally trained by minimizing the loss in eq.~\ref{MPC}
\begin{equation}\label{MPC}
    \mathscr{L}_{MPC} = \sum^{T}_{i=1} w_i \left \| x_{i} - y_{i} \right \|
\end{equation}
where $w$ weighs the contribution of masked and unmasked regions. \\
\textbf{Contrastive-learning} \cite{oord2018representation, baevski2020wav2vec} is a discriminative learning-based S3RL method. It maximizes the distance between the embeddings of positive and negative sample pairs. Same as \cite{baevski2020wav2vec}, we used vector quantization~\cite{jegou2010product} to determine the centroids of different anchor classes and used a weighted sum ($\beta$=0.1) between the info Noise-Contrastive Estimation (infoNCE)~\cite{oord2018representation} and a diversity loss ($\mathscr{L}_{D}$)~\cite{baevski2020wav2vec} as training objective.
\begin{equation}\label{CL}
    \mathscr{L}_{CL} = -\sum^{T}_{i=1} w_i\log \frac{\exp{(\Phi(y_i, q_i)} / \kappa)}{\sum_{\Tilde{q}\in Q_p}\exp{(\Phi(y_i, \Tilde{q})} / \kappa)} + \beta \mathscr{L}_{D}
\end{equation}
In eq.~\ref{CL}, $Q_p$ refers to the acoustic-unit level centroids that were quantized using the Gumbel softmax operation~\cite{jang2016categorical}; $q_i$ refer to the positive sample of $y_i$, calculated using the \textit{i-}th position in the original input. We consider the remaining codebook entries as negative samples and a cosine similarity distance function $\Phi(\cdot)$. As in MPC, $w$ is a function that weighs the contribution of masked and unmasked regions. $\kappa$ is a temperature scaling factor set to 0.1 as in~\cite{baevski2020wav2vec}.
$\mathscr{L}_{D}$ encourages the usage of different code entries and is defined as $\mathscr{L}_{D} = 1/GV \sum^{G}_{g=1}\sum^{V}_{v=1}p_{g,v}\log p_{g,v} $, where \textit{G} is the codebook number, \textit{V} is the entry size of each codebook, $p_{g,v}$ is the probability of selecting \textit{v-}th entry of the \textit{g-}th codebook, and is calculated using a softmax function across a batch of training sequences.
\vspace{-1.5mm}
\section{Proposed approach}
\label{sec:model}
\vspace{-1mm}
Here we introduce the small-footprint transformer model which was used in our experiments. Following that, we present our proposed method to enhance utterance-wise distinctions, which improves models' ability to differentiate utterances during S3RL.
\vspace{-2.5mm}
\subsection{Light-weight transformer}
\vspace{-1mm}
\label{ssec:light_transformer}
Since its introduction in~\cite{vaswani2017attention}, transformer and variants have found vast application in CV and NLP. The majority of these models exceed millions of parameters~\cite{devlin2018bert,baevski2020wav2vec,liu2020mockingjay,oord2018representation,jiang2019improving}, making them not suitable for KS on devices that are constrained by limited computational resources. Computation of the attention-matrix on transformers has time complexity \textit{$T^2$}, where \textit{$T$} refers to the input sequence length. Reducing the input size can mitigate computational demand. We accomplished this by processing the input with a VGG-like model~\cite{simonyan2014very} and by using a strided-convolution-based 2$\times$ down-sampling factor. To make our transformer mobile-friendly, we limited it to 330K parameters by decreasing both depth and width, making it comparable in size with KS models in \cite{khursheed2021tiny}. We implemented a number of simple, yet effective modifications to the adopted S3RL methods. In APC, models exploit bidirectional information hence violate the concept of auto-regressive methods where only past information should be used. We solved this by introducing an attention mask which made our model focus only on past timestamps. Compared to the original MPC approach where binary masks are randomly generated to prevent the model from exploiting local smoothness due to the continuous property of acoustic features~\cite{jiang2019improving,liu2020mockingjay}, we took inspiration from~\cite{baevski2020wav2vec} and used learnable mask-embeddings. We set weights to 0 and 1.0 for unmasked and masked region respectively, to focus on reconstructing missing features. 
\begin{table*}[!tb]
\resizebox{2\columnwidth}{!}{
\begin{threeparttable}
\caption{Experimental results obtained on Google Speech Command V2 and on the in-house keyword spotting datasets.}
\label{table:1}
\centering
\begin{tabular}{lcccccccccc}
 \hline\hline
 & \multicolumn{3}{c}{\textbf{Training strategy}} & \multicolumn{3}{c}{\textbf{Google speech commands v2 dataset}} & & \multicolumn{3}{c}{\textbf{In-house dataset}} \\
 \cline{2-3}\cline{5-7}\cline{9-11}
 \multirow{1}{*}{\textbf{Approach}} & \multirow{1}{*}{\textbf{PT}} & \multirow{1}{*}{\textbf{FT}} & & \multirow{1}{*}{\textbf{PT/FT}} & & \multirow{1}{*}{\textbf{Accuracy}} & &  \multirow{1}{*}{\textbf{PT/FT}} & & \multirow{1}{*}{\textbf{Relative FAR}} \\ 
 \hline
 Baseline$\_$MLP & $\times$ & Decoder only & & -/GC$\_$V2 && 39.2\% && -/- && - \\ 
 Baseline$\_$LT & $\times$ & Encoder+Decoder && -/GC$\_$V2 && 94.6\% && -/IH$\_$1.6K && 1.0 \\ 
 LT$\_$APC$_{mlp}$ & APC & Decoder only && LS$\_$960/GC$\_$V2 && 61.5\% && IH$\_$15K/IH$\_$1.6K && 1.371 \\
 LT$\_$APC & APC & Encoder+Decoder && LS$\_$960/GC$\_$V2 && 95.3\% && IH$\_$15K/IH$\_$1.6K && 0.827 \\
 LT$\_$MPC & MPC & Encoder+Decoder && LS$\_$960/GC$\_$V2 && 95.0\% && IH$\_$15K/IH$\_$1.6K && 0.838 \\
 LT$\_$CL & CL & Encoder+Decoder && LS$\_$960/GC$\_$V2 && 94.9\% && IH$\_$15K/IH$\_$1.6K && 0.854 \\
 \textbf{LT$\_$APC+} & APC$\_$uwdb & Encoder+Decoder && LS$\_$960/GC$\_$V2 && \textbf{95.8\%} && IH$\_$15K/IH$\_$1.6K && \textbf{0.818} \\
 LT$\_$MPC+ & MPC$\_$uwdb & Encoder+Decoder && LS$\_$960/GC$\_$V2 && 95.3\% && IH$\_$15K/IH$\_$1.6K && 0.831 \\
 LT$\_$CL+ & CL$\_$uwdb  & Encoder+Decoder && LS$\_$960/GC$\_$V2 && 95.3\% && IH$\_$15K/IH$\_$1.6K && 0.847 \\
 \hline\hline
\end{tabular}
\begin{tablenotes}
\item[Legend] \textbf{GC$\_$V2}=Google Speech Command V2 dataset; \textbf{LS$\_$960}=Librispeech 960hrs dataset; \textbf{IH$\_$15K}=in-house 15K hours of KS audio used as unannotated data; \textbf{IH$\_$1.6K}=in-house 1.6K hours of annotated KW$_1$ audio sub-dataset. \textbf{uwdb}=utterance-wise distinction boosting; \textbf{Relative FAR}=relative false acceptance rate compared to baseline at fixed false rejection rate; \textbf{PT/FT}=pre-training/fine-tuning; \textbf{Decoder}=one MLP layer only; \textbf{Encoder}= Transformer blocks.
 \vspace{-3.5mm}
\end{tablenotes}
\end{threeparttable}
}
\end{table*}
\vspace{-2.5mm}
\subsection{Improving utterance-wise distinction}
\vspace{-1mm}
\cite{chen2022unispeech} used CL to enhance model's ability to learn different speakers characteristics. This speaker-wise CL provided a performance boost on down-stream tasks including speaker identification and keyword-spotting. While \cite{chen2022unispeech} focused on speaker-difference, we are interested in learning general and abstract differences among utterances. Similar to past research~\cite{yang2021superb,gong2022ssast,chen2022unispeech}, we captured utterance level differences by processing the S3RL-learned features with a mean-pooling operation followed by a linear transformation. In addition, we propose a two-step mechanism to enhance utterance-wise distinction. Our approach is schematically depicted in Fig.~\ref{figure utterance-wise distinction boosting}. The method comprises two blocks: the first block is a general S3RL block, which includes a light-weight transformer ($LWT_{1}$). $LWT_{1}$ can be trained using any S3RL method. Inputs are pre-processed as described in section \ref{ssec:s3rl} to accommodate specific S3RL input requirements. The second block includes another transformer model ($LWT_{2}$) pre-trained with the same S3RL method as $LWT_{1}$. Weights of $LWT_{2}$ are frozen, so that $LWT_{2}$ is used as a feature extractor. Inputs to $LWT_{2}$ are not pre-processed to obtain the utterance-level representation. The training procedure is as follows: \textit{i)} intermediate output features are extracted at the second layer of $LWT_{1}$ and $LWT_{2}$, \textit{ii)} these features are processed by a mean-pooling operation to generate utterance-level representation $u_{1}$ for $LWT_{1}$ and $u_{2}$ for $LWT_{2}$ and \textit{iii)} $u_{2}$ representations are fed into an anchoring block to generate positive ($\Tilde{Q}\textsuperscript{+}$) and negative ($\Tilde{Q}\textsuperscript{-}$) sample pairs based on vector quantization. 
With regard to $LWT_{1}$, weights are tuned in a multi-task learning fashion, simultaneously solving S3RL and utterance-wise contrastive learning. For utterance-wise CL, we minimize a combination of the infoNCE contrastive learning loss function ($\mathscr{L}_{utt}$) and the diversity loss $\mathscr{L}_{D}$ with $\beta$=0.1:
\begin{equation}\label{contrastive loss}
    \mathscr{L}_{utt} = -\log \frac{\exp{(\Phi(u_1, \Tilde{Q}\textsuperscript{+})} / \kappa)}{\sum_{\Tilde{q}\in Q}\exp{(\Phi(u_1, \Tilde{q})} / \kappa)} + \beta \mathscr{L}_{D}
\end{equation}
The complete training objective function used in our experiments is reported in eq.~\ref{final equation} and to maintain self-supervised learning properties, we set $\alpha$ to 0.9.
\begin{align}\label{final equation}
    \mathscr{L}=\alpha \mathscr{L}_{S3RL} + (1-\alpha) \mathscr{L}_{utt}
\end{align}
\vspace{-1.5mm}
\section{Experimental study}
\vspace{-1mm}
\vspace{-2.5mm}
\subsection{Datasets}
\vspace{-1mm}
\label{ssec:datasets}
We used three datasets throughout our experimentation: one is comprised of de-identified audio files from our production data and the others are open-source datasets, widely used among the speech-processing research community. All data were processed in the form of 64-D LFBE spectrogram. These were formulated by using an analysis window and shift-size of 25ms and 10ms respectively. 
\textbf{Librispeech dataset} This dataset contains speech in English, sampled at 16kHz. We used training subsets of Librispeech~\cite{panayotov2015librispeech} for self-supervised pre-training, creating 960 hours of training corpus. \\
\textbf{Google speech commands v2 dataset} This dataset is comprised of 105,829 one second-long utterances of 35 keywords~\cite{warden2018speech} and was used for KS down-streaming task. \\
\textbf{In-house keyword spotting dataset} We collected 20K hours worth of de-identified audio containing four keywords (16.6K hours with KW$_1$, 1.6K hours with KW$_2$, 833 hours with KW$_3$ and 1K hours with KW$_4$), recorded at different front-end conditions for training purpose. We used KW$_1$ for single keyword experiments, where we split the datasets into 15K/1.6K hours for pre-training and fine-tuning, respectively. In multi-wakeword experiments, we used the entire dataset. Our collected test data consists of 8.5K hours with KW$_1$, 1K hours with KW$_2$, 444 hours with KW$_3$ and 500 hours with KW$_4$ audio streams. The keywords were semi-automatically annotated and inspected for quality check.
%
\vspace{-2.5mm}
\subsection{Training setup}
\vspace{-1mm}
\label{ssec:configurations}
\textbf{S3RL configuration} For APC, we set the number of predicted future frames $n$=8. This is equivalent to $n$=5 in~\cite{chung2020generative}, since the light VGG-style model which we used has a receptive-field of 7 frames. With respect to the MPC masking operation, we masked out 50\% of the frames and replaced them with mask-embedding, unlike \cite{jiang2019improving,liu2020mockingjay}. With regard to CL, to accommodate our limited model capacity, we adopted a single codebook with 64 entries. Similarly, for utterance-wise CL, we used a single codebook with 32 entries. \\
\textbf{Training configuration} The S3RL training objectives were defined in section~\ref{ssec:s3rl}. As shown in Fig 1, we applied three transformer blocks as the encoder and one MLP layer (projection) as the decoder. To fine-tune our models on the down-stream KS task, we minimized the cross-entropy loss function. We compared the performance with freezing and unfreezing setups as in \cite{yang2021superb} and \cite{devlin2018bert} respectively. Due to the model's limited capacity, we used the output from final layers as representations instead of a weighted sum of output from different layers~\cite{yang2021superb,chen2022wavlm,chen2022unispeech}. We did not apply SpecAug ~\cite{park2019specaugment} as this could obscure the the effectiveness of S3RL and add additional variables to account for. We ran pre-training for 20 epochs and fine-tuned our models for 10 additional epochs on the KS task, using Adam optimize with an initial learning rate set to 1e-3 and per-epoch exponential weight decay of rate 0.95. Each epoch consisted of 5000 steps with batch size equal to 2000 distributed across 8 NVIDIA Tesla V100 GPUs.
\vspace{-1.5mm}
\section{Results and discussion}
\vspace{-1mm}
\label{sec:res_and_dis}
\subsection{Google speech commands v2 dataset}\label{sec:gsc_v2}
\vspace{-1mm}
\textbf{S3RL is feasible in light-weight transformers} In table~\ref{table:1}, we compare model performance at various settings, in terms of accuracy, as common practice in the literature. We found that self-supervised pre-training with model fine-tuning achieved better performance than training from scratch (Baseline$\_$LT), irrespective of the S3RL method. This demonstrates S3RL is feasible on KS with tiny footprint models, using large amounts of unlabeled data. Furthermore, we observed LT$\_$APC$_{mlp}$ outperformed the baseline multi-layer perceptron (MLP) model (Baseline$\_$MLP). This demonstrates S3RL facilitates training light-weight models. Freezing the entire encoder and fine-tuning the decoder's weights (LT$\_$APC$_{mlp}$) degraded performance, compared to training from scratch (Baseline$\_$LT). We suppose this is because the model which we used for pre-training was not large enough to be able to learn general speech representations. The performance improvement of LT$\_$APC compared to Baseline$\_$LT indicates that down-stream tasks can benefit from pre-training, as it provides a better network initialization. From table~\ref{table:1}, the LT$\_$APC provided the largest performance improvement (0.7\% accuracy) among the remaining S3RL methods.\\
\textbf{Improving utterance-wise distinction positively impacts KS}
We conducted experiments using our proposed method to improve utterance-wise representations. We used pre-trained APC, MPC and CL models and denote these configurations with ``+". As reported in table~\ref{table:1}, APC+, MPC+ and CL+ outperformed original light-weight S3RL methods respectively. Specifically, LT$\_$APC+ provided 0.6\% accuracy improvement compared to LT$\_$APC and 1.2\% accuracy improvement compared to Baseline$\_$LT. We suggest this demonstrates the benefits of the proposed second-step utterance-wise distinction boosting. Intuitively, if we used larger utterance-wise distinction models for S3RL boosting, utterance level representations would improve even further. We leave this investigation for future work.
\begin{table}[tb]
\centering
\caption{KS results of the ablation study conducted on an in-house dataset. Performance is reported in terms of relative false acceptance rate (FAR) to the baseline model at fixed false rejection rate (FRR).}
\label{table:2}
\begin{tabular}{l c c c c c} 
 \hline\hline
 & \multicolumn{4}{c}{Fine-tuning dataset (hours)} \\
 \cline{2-5}
 Approach & 166 & 416 & 833 & 1.6K \\ 
 \hline
  Baseline$\_$LT & 1.35 & 1.27 & 1.09 & 1.0\\ 
  LT$\_$MPC+ & 1.20 & 1.12 & \textbf{0.901} & 0.831 \\
  LT$\_$CL+ & 1.31 & 1.12 & \textbf{0.903} & 0.847 \\
  \textbf{LT$\_$APC+} & 1.19 & \textbf{1.07} & \textbf{0.895} & 0.818 \\
 \hline
 & \multicolumn{4}{c}{$n$ future in APC} \\
 \cline{2-5}
 Approach & 5 & 8 & 10 & 20 \\
 \hline
  LT$\_$APC+ & 0.910 & \textbf{0.818} & 0.852 & 0.973 \\
 \hline
 & \multicolumn{4}{c}{Mask proportion in MPC} \\
 \cline{2-5}
 Approach & 15\% & 25\% & 50\% & 75\% \\
 \hline
  LT$\_$MPC+ & 0.953 & 0.878 & \textbf{0.831} & 0.950 \\
 \hline
 & \multicolumn{4}{c}{Codebook size in CL} \\
 \cline{2-5}
 Approach & 32 & 64 & 128 & 256 \\
 \hline
  LT$\_$CL+ & 0.872 & \textbf{0.847} & 0.859 & 0.862 \\
 \hline
 \hline
\end{tabular}
\vspace{-3.5mm}
\end{table}
\vspace{-2.5mm}
\subsection{In-house keyword spotting dataset}
\vspace{-1mm}
\label{sec:in_house_ks}
We conducted performance analysis on our in-house dataset by means of false acceptance rate (FAR) relative to the baseline model at fixed false rejection rate (FRR). For a specific keyword, the FRR is the ratio between false negative and true positive samples at the operating point (OP) of the baseline model. We identified an OP at which the proposed approach showed similar FRR and used that OP to calculate the corresponding FAR, which is the ratio between false positives and true negatives. As reported in table~\ref{table:1}, APC and APC+ achieved better performance among S3RL methods. The LT$\_$APC showed 17.3\% relative FAR improvement (from 1.0 to 0.827) compared to Baseline$\_$LT. The LT$\_$APC+ further reduced the relative FAR to 0.818 from 0.827. These results demonstrate that S3RL can be effective even when using light-weight models and large scale datasets. On the other hand, these results show that CL-based S3RL did not perform on-par with the other S3RL methods. We speculate this is due to the limited model capacity, which constrained the model learning capability. We further conducted an ablation study on the same in-house KW$_1$ dataset and report results in table~\ref{table:2}. When we fine-tuned with just 50\% of the data, we performed on-par or even outperformed the model trained from scratch (Baseline$\_$LT). In our ablation study, using 1.6K hours worth of KW$_1$ training audio for fine-tuning, APC, MPC and CL based approaches achieved best performance with future shift $n$=8, 50\% mask proportion and 64 entries in the codebook. We extended the experiment to multiple keywords from our in-house dataset. Instead of using a single keyword during pre-training, we combined the first three keyword data for pre-training and left the fourth keyword out as an independent dataset to prove the out of domain effectiveness in S3RL. We fine-tuned the model for each keyword independently using the corresponding audio samples. Results in table~\ref{table:3} show that the LT$\_$APC+ performed better than the Baseline$\_$LT on all four keywords task. It achieved 12,4\% and 23.7\% relative FAR improvement on KW$_2$ and KW$_3$. Even without in-domain data in the pre-training dataset, the model still achieved a 6\% relative FAR gain on KW$_4$. These results indicate that when data are scarce, down-stream tasks can benefit from S3RL. 
\vspace{-2.5mm}
\begin{table}[tb]
\centering
\caption{Multi-keyword spotting results on an in-house dataset. Performance is reported in terms of relative false acceptance rate (FAR) to the baseline model at fixed false rejection rate (FRR).}
\label{table:3}
\begin{tabular}{c c c c c} 
 \hline\hline
 Approach  & KW$_1$ & KW$_2$ & KW$_3$ & KW$_4$\\ 
 \hline
 Training dataset (hours) & 16.6K & 1.6K  & 833  & 1.1K \\ 
 \hline
  Baseline$\_$LT & 1.0 & 1.0 & 1.0 & 1.0\\ 
  LT$\_$MPC+ & \textbf{0.906 }& 0.915 & 0.878 & 0.960 \\
  LT$\_$CL+ & 0.919 & 0.920 & 0.886 & 0.957 \\
  \textbf{LT$\_$APC+} & 0.910 & \textbf{0.876} & \textbf{0.763} & \textbf{0.940} \\ 
 \hline\hline
\end{tabular}
\vspace{-3.5mm}
\end{table}
\vspace{-1.0mm}
\section{Conclusion and future work}
\vspace{-1mm}
\label{sec:conclusion_and_future}
In this paper, we explored the feasibility of using self-supervised representation learning on small-footprint models. Specifically, we pre-trained light-transformers by means of three S3RL methods, such as APC, MPC and CL and subsequently fine-tuned the transformers on keyword-spotting down-stream tasks. In our experimental study, all three S3RL methods showed better performance compared to training from scratch. Pre-training using APC provided the best results on both public and in-house datasets. In addition, we proposed a method that combines S3RL and CL to boost representation learning by extracting more diverse utterance-level representations. We showed that the proposed combination led to better performance on the down-stream keyword-spotting task. It is important to remark that our method can be easily implemented and added to any S3RL methods. As part of the future work, additional speech representation learning methods should be investigated, as well as alternative large models to extract utterance-level representation. It would be interesting to explore the effect of learning representations by means of a model as proposed by~\cite{stafylakis2022extracting}, as opposed to the mean pooling approach which we adopted in this work. 

\bibliographystyle{IEEEbib}
\bibliography{refs}

\begin{thebibliography}{10}

\bibitem{devlin2018bert}
Jacob Devlin, Ming-Wei Chang, Kenton Lee, and Kristina Toutanova,
\newblock ``Bert: Pre-training of deep bidirectional transformers for language
  understanding,''
\newblock in {\em NAACL}, 2019.

\bibitem{peters-etal-2018-deep}
Matthew~E. Peters, Mark Neumann, Mohit Iyyer, Matt Gardner, Christopher Clark,
  Kenton Lee, and Luke Zettlemoyer,
\newblock ``Deep contextualized word representations,''
\newblock in {\em ACL}, 2018.

\bibitem{he2020momentum}
Kaiming He, Haoqi Fan, Yuxin Wu, Saining Xie, and Ross Girshick,
\newblock ``Momentum contrast for unsupervised visual representation
  learning,''
\newblock in {\em CVPR}, 2020.

\bibitem{he2022masked}
Kaiming He, Xinlei Chen, Saining Xie, Yanghao Li, Piotr Doll{\'a}r, and Ross
  Girshick,
\newblock ``Masked autoencoders are scalable vision learners,''
\newblock in {\em CVPR}, 2022.

\bibitem{yang2021superb}
Shu-wen Yang, Po-Han Chi, Yung-Sung Chuang, Cheng-I~Jeff Lai, Kushal Lakhotia,
  Yist~Y Lin, Andy~T Liu, Jiatong Shi, Xuankai Chang, Guan-Ting Lin, et~al.,
\newblock ``Superb: Speech processing universal performance benchmark,''
\newblock in {\em arXiv:2105.01051}, 2021.

\bibitem{chung2020generative}
Yu-An Chung and James Glass,
\newblock ``Generative pre-training for speech with autoregressive predictive
  coding,''
\newblock in {\em ICASSP}, 2020.

\bibitem{jiang2019improving}
Dongwei Jiang, Xiaoning Lei, Wubo Li, Ne~Luo, Yuxuan Hu, Wei Zou, and Xiangang
  Li,
\newblock ``Improving transformer-based speech recognition using unsupervised
  pre-training,''
\newblock in {\em arXiv:1910.09932}, 2019.

\bibitem{liu2020mockingjay}
Andy~T Liu, Shu-wen Yang, Po-Han Chi, Po-chun Hsu, and Hung-yi Lee,
\newblock ``Mockingjay: Unsupervised speech representation learning with deep
  bidirectional transformer encoders,''
\newblock in {\em ICASSP}, 2020.

\bibitem{hsu2021hubert}
Wei-Ning Hsu, Benjamin Bolte, Yao-Hung~Hubert Tsai, Kushal Lakhotia, Ruslan
  Salakhutdinov, and Abdelrahman Mohamed,
\newblock ``Hubert: Self-supervised speech representation learning by masked
  prediction of hidden units,''
\newblock in {\em IEEE/ACM Transactions on Audio, Speech, and Language
  Processing}, 2021.

\bibitem{chang2022distilhubert}
Heng-Jui Chang, Shu-wen Yang, and Hung-yi Lee,
\newblock ``Distilhubert: Speech representation learning by layer-wise
  distillation of hidden-unit bert,''
\newblock in {\em ICASSP}, 2022.

\bibitem{oord2018representation}
Aaron van~den Oord, Yazhe Li, and Oriol Vinyals,
\newblock ``Representation learning with contrastive predictive coding,''
\newblock in {\em NeurIPS}, 2018.

\bibitem{ling2020deep}
Shaoshi Ling, Yuzong Liu, Julian Salazar, and Katrin Kirchhoff,
\newblock ``Deep contextualized acoustic representations for semi-supervised
  speech recognition,''
\newblock in {\em ICASSP}, 2020.

\bibitem{baevski2020wav2vec}
Alexei Baevski, Yuhao Zhou, Abdelrahman Mohamed, and Michael Auli,
\newblock ``wav2vec 2.0: A framework for self-supervised learning of speech
  representations,''
\newblock in {\em NeurIPS}, 2020.

\bibitem{ling2020decoar}
Shaoshi Ling and Yuzong Liu,
\newblock ``Decoar 2.0: Deep contextualized acoustic representations with
  vector quantization,''
\newblock in {\em arXiv:2012.06659}, 2020.

\bibitem{gong2022ssast}
Yuan Gong, Cheng-I Lai, Yu-An Chung, and James Glass,
\newblock ``Ssast: Self-supervised audio spectrogram transformer,''
\newblock in {\em AAAI}, 2022.

\bibitem{bovbjerg2022improving}
Holger~Severin Bovbjerg and Zheng-Hua Tan,
\newblock ``Improving label-deficient keyword spotting using self-supervised
  pretraining,''
\newblock in {\em arXiv:2210.01703}, 2022.

\bibitem{baevski2022data2vec}
Alexei Baevski, Wei-Ning Hsu, Qiantong Xu, Arun Babu, Jiatao Gu, and Michael
  Auli,
\newblock ``Data2vec: A general framework for self-supervised learning in
  speech, vision and language,''
\newblock in {\em arXiv:2202.03555}, 2022.

\bibitem{warden2018speech}
Pete Warden,
\newblock ``Speech commands: A dataset for limited-vocabulary speech
  recognition,''
\newblock in {\em arXiv:1804.03209}, 2018.

\bibitem{jegou2010product}
Herve Jegou, Matthijs Douze, and Cordelia Schmid,
\newblock ``Product quantization for nearest neighbor search,''
\newblock in {\em IEEE PAMI}, 2010.

\bibitem{jang2016categorical}
Eric Jang, Shixiang Gu, and Ben Poole,
\newblock ``Categorical reparameterization with gumbel-softmax,''
\newblock in {\em ICLR}, 2017.

\bibitem{vaswani2017attention}
Ashish Vaswani, Noam Shazeer, Niki Parmar, Jakob Uszkoreit, Llion Jones,
  Aidan~N Gomez, {\L}ukasz Kaiser, and Illia Polosukhin,
\newblock ``Attention is all you need,''
\newblock in {\em NeurIPS}, 2017.

\bibitem{simonyan2014very}
Karen Simonyan and Andrew Zisserman,
\newblock ``Very deep convolutional networks for large-scale image
  recognition,''
\newblock in {\em arXiv:1409.1556}, 2014.

\bibitem{khursheed2021tiny}
Mohammad~Omar Khursheed, Christin Jose, Rajath Kumar, Gengshen Fu, Brian Kulis,
  and Santosh~Kumar Cheekatmalla,
\newblock ``Tiny-crnn: Streaming wakeword detection in a low footprint
  setting,''
\newblock in {\em ASRU}, 2021.

\bibitem{chen2022unispeech}
Sanyuan Chen, Yu~Wu, Chengyi Wang, Zhengyang Chen, Zhuo Chen, Shujie Liu, Jian
  Wu, Yao Qian, Furu Wei, Jinyu Li, et~al.,
\newblock ``Unispeech-sat: Universal speech representation learning with
  speaker aware pre-training,''
\newblock in {\em ICASSP}, 2022.

\bibitem{panayotov2015librispeech}
Vassil Panayotov, Guoguo Chen, Daniel Povey, and Sanjeev Khudanpur,
\newblock ``Librispeech: an asr corpus based on public domain audio books,''
\newblock in {\em ICASSP}, 2015.

\bibitem{chen2022wavlm}
Sanyuan Chen, Chengyi Wang, Zhengyang Chen, Yu~Wu, Shujie Liu, Zhuo Chen, Jinyu
  Li, Naoyuki Kanda, Takuya Yoshioka, Xiong Xiao, et~al.,
\newblock ``Wavlm: Large-scale self-supervised pre-training for full stack
  speech processing,''
\newblock in {\em IEEE Journal of Selected Topics in Signal Processing}, 2022.

\bibitem{park2019specaugment}
Daniel~S Park, William Chan, Yu~Zhang, Chung-Cheng Chiu, Barret Zoph, Ekin~D
  Cubuk, and Quoc~V Le,
\newblock ``Specaugment: A simple data augmentation method for automatic speech
  recognition,''
\newblock in {\em arXiv:1904.08779}, 2019.

\bibitem{stafylakis2022extracting}
Themos Stafylakis, Ladislav Mosner, Sofoklis Kakouros, Oldrich Plchot, Lukas
  Burget, and Jan Cernocky,
\newblock ``Extracting speaker and emotion information from self-supervised
  speech models via channel-wise correlations,''
\newblock in {\em arXiv:2210.09513}, 2022.

\end{thebibliography}


\end{document}